\documentclass[twocolumn,showpacs,preprintnumbers,amsmath,amssymb]{revtex4}

\usepackage{graphicx}
\usepackage{dcolumn}
\usepackage{bm}
\newcommand{\figref}[1]{Figure~\ref{#1}}
\newcommand{\remove}[1]{} 

\begin{document}

\preprint{APS/123-QED}

\title{Parameterized Centrality for Network Analysis}

\author{Kristina Lerman}
 \email{lerman@isi.edu}
\author{Rumi Ghosh}%
 \email{rumig@usc.edu}
\affiliation{%
USC Information Sciences Institute\\
4676 Admiralty Way, Marina del Rey, CA 90292
}%

\date{\today}

\begin{abstract}
Bonacich centrality measures the number of attenuated paths between nodes in a network. We use this metric to study network structure, specifically, to rank nodes and find community structure of the network. To this end we extend the modularity-maximization method for community detection to use this centrality metric as a measure of node connectivity. Bonacich centrality contains a tunable parameter that sets the length scale of interactions. By studying how rankings and discovered communities change when this parameter is varied allows us to identify globally important nodes and structures.
We apply the proposed method to several benchmark networks and show that it leads to better insight into network structure than earlier methods.
\end{abstract}

\pacs{89.75.Hc, 89.20.Hh, 89.65.Ef, 02.10.Ud}
\maketitle

Centrality measures the degree to which network structure determines importance of a node in a network.
Over the years many different centrality metrics have been studied.
Katz~\cite{Katz} recognized that an individual's centrality depends not only on how many others she is connected to (her degree), but also on their centrality. He  measured centrality of a node by the total number of paths linking it to other nodes in a network, exponentially weighted by the length of the path. Freeman~\cite{Freeman} defined betweenness centrality as the fraction of all shortest paths between pairs of nodes that pass through a given node. Several variants of centrality based on random walks have been proposed and analyzed~\cite{Bonacich87,Stephenson89,PageRank,Noh02,Newman05}.
Specifically, Bonacich~\cite{Bonacich87}, like Katz, measured the total number of attenuated paths from a node, but now the attenuation factors along direct (from the originating node) and indirect (from intermediate nodes) edges in a path can be different. These parameters set the length scale of interactions. Unlike other centrality metrics, which do not distinguish between local and global structure, these parameterized centrality metrics can differentiate between locally connected nodes, i.e., nodes that are linked to interconnected nodes, and globally connected nodes, which are linked to and mediate communication between otherwise unconnected nodes.

In addition to ranking nodes, Bonacich centrality can be used to identify communities within the network. In this paper, we generalize modularity maximization-based approach~\cite{Newman104,Newman206}  to use Bonacich centrality. Rather than find regions of the network that have greater than expected number of edges connecting nodes~\cite{GirvanNewman04}, our approach looks for regions that have greater than expected number of paths between nodes.
Arenas et al.~\cite{Arenas} have similarly generalized modularity to find correlations between nodes beyond nearest neighbors. Their motif-based community detection algorithm uses the size of the motif to impose a limit on the proximity of neighbors. Our method, on the other hand, imposes no such limit. The measure of global correlation computed using Bonacich centrality  is equal to the weighted average of correlations  for motifs of different sizes. Our method enables us to easily calculate this complex term.

We use Bonacich centrality to study the structure of several benchmark networks, as well as the network extracted from a social web site. We show that parameterized centrality can identify locally and globally important nodes and structures, leading to a better understanding of network structure.

Bonacich~\cite{Bonacich87} defined a centrality metric $C_{i,j}(\alpha,\beta)$ as the total number of attenuated paths between nodes $i$ and $j$, with $\beta$ and $\alpha$ giving the attenuation factors along direct edges (from $i$) and indirect edges (from intermediate nodes) in the path from $i$ to $j$.
Given the adjacency matrix of the network $A$, Bonacich centrality matrix can be computed as follows:
\begin{equation}
\label{eq:inf}
C(\alpha,\beta) = \beta A +\beta \alpha_1 A \cdot A + \cdots + \beta \prod_{j=1}^n \alpha_j  A^{n+1} + \cdots
\end{equation}
\noindent
The first term gives the number of paths of length one (edges) from $i$ to $j$, the second the number of paths of length two, etc. Although $\alpha_j$ along different edges in a path could in principle be different, for simplicity, we take them all to be equal: $\alpha_j=\alpha$ for all $j$. In this case, the series converges to $C(\alpha,\beta)=\beta A {( I-\alpha A)}^{-1}$, which holds while  $\alpha < 1/\lambda$, where $\lambda$ is the largest characteristic root of  $A$~\cite{Ferrar}.  For $\alpha=\beta$, Bonacich centrality reduces to the Katz score~\cite{Katz}.

Bonacich centrality (b-centrality) contains a tunable parameter $\alpha$ that sets the length scale of interactions.
For $\alpha=0$ (and $\beta=1$), b-centrality takes into account direct edges only and reduces to degree centrality. As $\alpha$ increases, $C(\alpha,\beta)$ becomes a more global measure, taking into account ever larger network components. The expected length of a path, the radius of centrality, is $(1-\alpha)^{-1}$. This tunable parameter turns b-centrality into a powerful tool for studying network structure.

Following Bonacich, we use  $C_i(\alpha,\beta)=\sum_j{C_{ij}(\alpha,\beta)}$ as the measure of how `close' node $i$ is to other nodes in a network. A node has high centrality if it is connected to many highly interconnected nodes, i.e., it is a \emph{leader} within its community. A node can also have high centrality if it is connected to nodes from different communities. Such \emph{mediators} bridge different communities, enabling communication between them~\cite{Granovetter}. We can identify such nodes because their b-centrality increases with $\alpha$. Other centrality metrics do not distinguish between locally and globally connected nodes.

Girvan \& Newman~\cite{GirvanNewman04} proposed \emph{modularity} as a metric for evaluating community structure of a network.
The modularity-optimization class of community detection algorithms~\cite{Newman104,Newman204,Newman206} finds a network division that maximizes the modularity $Q=$(connectivity within community)-(expected connectivity), where connectivity is density of edges.
We extend this definition to use b-centrality as the measure of network connectivity~\cite{Ghosh08}.
Therefore, in the best division of a network, nodes have more paths connecting them to nodes within their community than to outside nodes.
We generalize modularity $Q$ as
\begin{equation}
\label{eq: mod2}
Q(\alpha)=\sum_{ij} {[C_{ij} - \bar{C}_{ij}]\delta(s_i, s_j)}
\end{equation}
where $C_{ij}$ is given by Eq.~\ref{eq:inf}, $\bar{C}_{ij}$ is the expected b-centrality, and $s_i$ is the index of the community $i$ belongs to, with $\delta(s_i, s_j) = 1$ if $s_i =s_j$; otherwise, $\delta(s_i, s_j)=0$. We round the values of $C_{ij}$ to the nearest integer. Since $\beta$ factors out of modularity, we consider dependence on $\alpha$ only.

To compute expected centrality, we consider a graph, referred to as the null model, which has the same number of nodes and edges as the original graph, but in which the edges are placed at random. To make the derivation below more intuitive, instead of b-centrality we talk of the number of attenuated paths.
When all the nodes are placed in a single group, then  axiomatically, $Q=0$. Therefore $ \sum_{ij}[C_{ij} - \bar{C}_{ij}] =0$, and we set
$W = \sum_{ij} \bar{C}_{ij}=\sum_{ij} C_{ij}.$
Therefore, according to the argument above, the total number of paths between nodes in the null model $\sum_{ij} \bar{C}_{ij}$ is equal to the total number of paths in the original graph, $\sum_{ij} C_{ij}$.  We further restrict the choice of null model to one where the expected number of paths reaching node $j$, $W_j^{in}$, is equal to the actual number of paths reaching the corresponding node in the original graph.
$
W_{j}^{in} = \sum_{i} \bar{C}_{ij} = \sum_{i} C_{ij}\,$.
Similarly, we also assume that in the null model, the expected number of paths originating at node $i$, $W_{i}^{out}$, is equal to the actual number of paths originating at the corresponding node in the original graph
$
W_{i}^{out} = \sum_{j} \bar{C}_{ij} = \sum_{j} C_{ij}\,.
$
Next, we reduce the original graph $G$ to a new graph $G^{\prime}$ that has the same number of nodes as $G$  and total number of edges  $W$, such that each edge has weight 1 and the number of edges between nodes $i$ and $j$ in $G^{\prime}$ is  $C_{ij}$. Now the expected number of paths between  $i$ and $j$  in graph $G$ could be taken as the expected number of the edges between nodes $i$ and $j$ in graph $G^{\prime}$, and the actual number of paths between nodes $i$ and $j$  in graph $G$ can be taken as the actual number of edges  between node $i$ and node $j$ in graph $G^{\prime}$. The equivalent random graph $G''$ is used to find the \emph{expected}  number of edges  from node $i$ to node $j$. In this graph  the edges are placed in random subject to constraints:
(\emph{i}) The total number of edges in $G''$ is $W$;
(\emph{ii}) The out-degree of node $i$ in $G''$ = out-degree of node $i$ in $G^{\prime} = W_{i}^{out}$;
(\emph{iii}) The in-degree of a node $j$ in graph $G''$ =in-degree of node $j$ in graph $G^{\prime} =W_{j}^{in}$.
Thus in $G''$ the  probability that an edge will emanate from a particular node  depends only on the out-degree of that node; the probability that an edge is incident on a particular node depends only on the  in-degree of that node; and the probabilities of the two nodes being the two ends of a single edge are independent of each other. In this case, the probability that an edge exists from $i$ to $j$ is given by $C$(\emph{emanates from i}) $\cdot $ $C$(\emph{incident on j})=$(W_{i}^{out}/W)(W_{j}^{in}/W)$.
Since the total number of edges is $W$ in $G''$, therefore the expected number of edges between $i$ and $j$ is $W \cdot (W_{i}^{out}/W)(W_{j}^{in}/W)=\bar{C}_{ij}$, the expected the expected b-centrality in $G$.

Once we compute $Q(\alpha)$, we have to select an algorithm to divide the network into communities that optimize $Q(\alpha)$. Brandes et al.~\cite{Brandes} have shown that the decision version of modularity maximization is NP-complete. Like others~\cite{Newman206,Leicht}, we use the leading eigenvector method to obtain an approximate solution. In this method, nodes are assigned to either of two groups based on a single eigenvector  corresponding to the largest  positive eigenvalue of the modularity matrix. This process is repeated for each group until modularity does not increase further upon division.


We apply the formalism developed above to benchmark networks studied in literature, and a network extracted from the social photosharing site Flickr.
We use \emph{purity} to evaluate the quality of discovered communities. We define purity as the fraction of all pairs of objects in the same community that are assigned to the same group by the algorithm. This is a simplified version of the Wallace criterion~\cite{Wallace} for evaluating performance of clustering algorithms.

\begin{figure}[tbh]
\begin{tabular}{c}
     \includegraphics[width=0.33\textwidth]{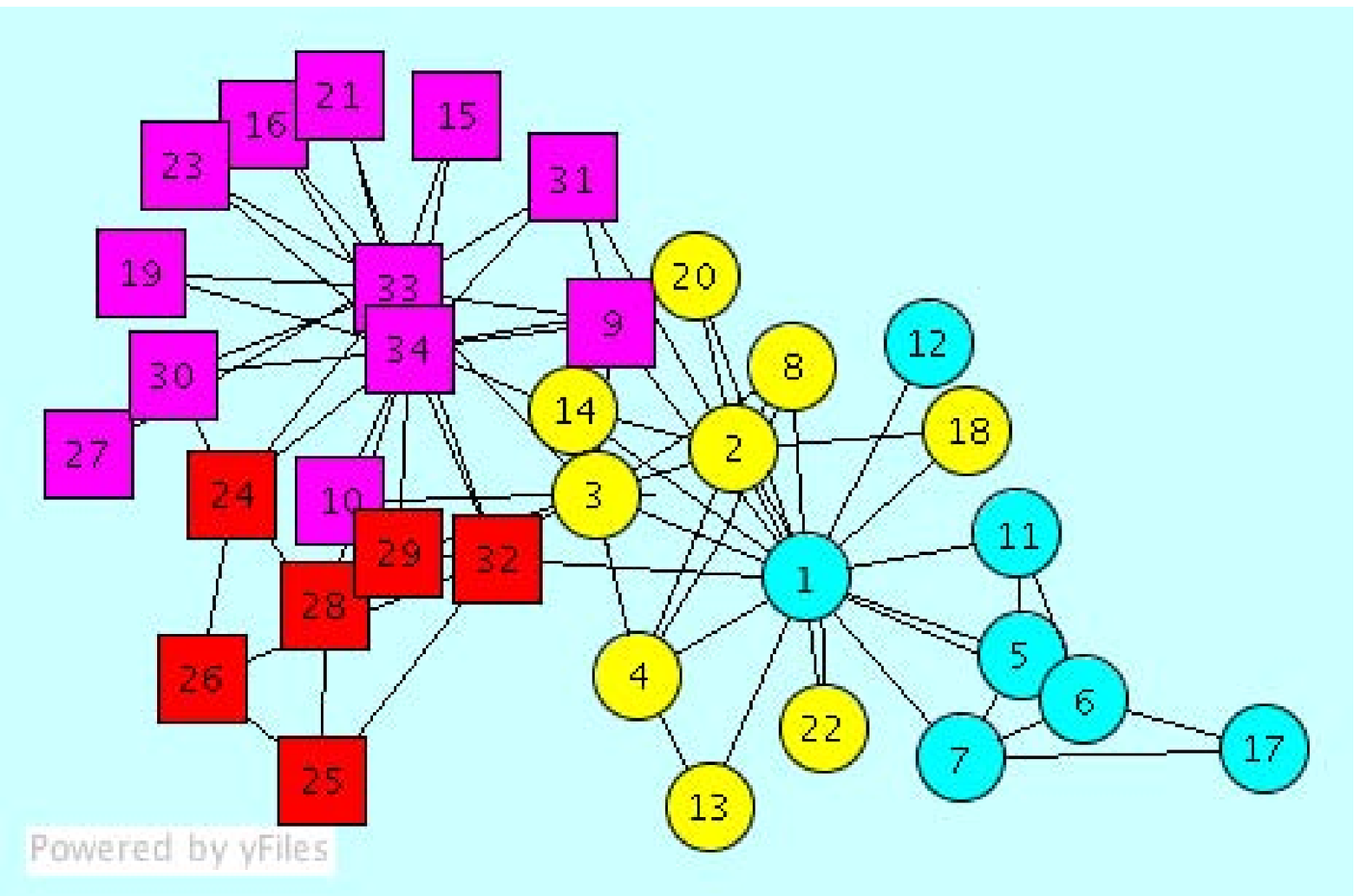} \\
     (a) \\
 \includegraphics[width=0.4\textwidth]{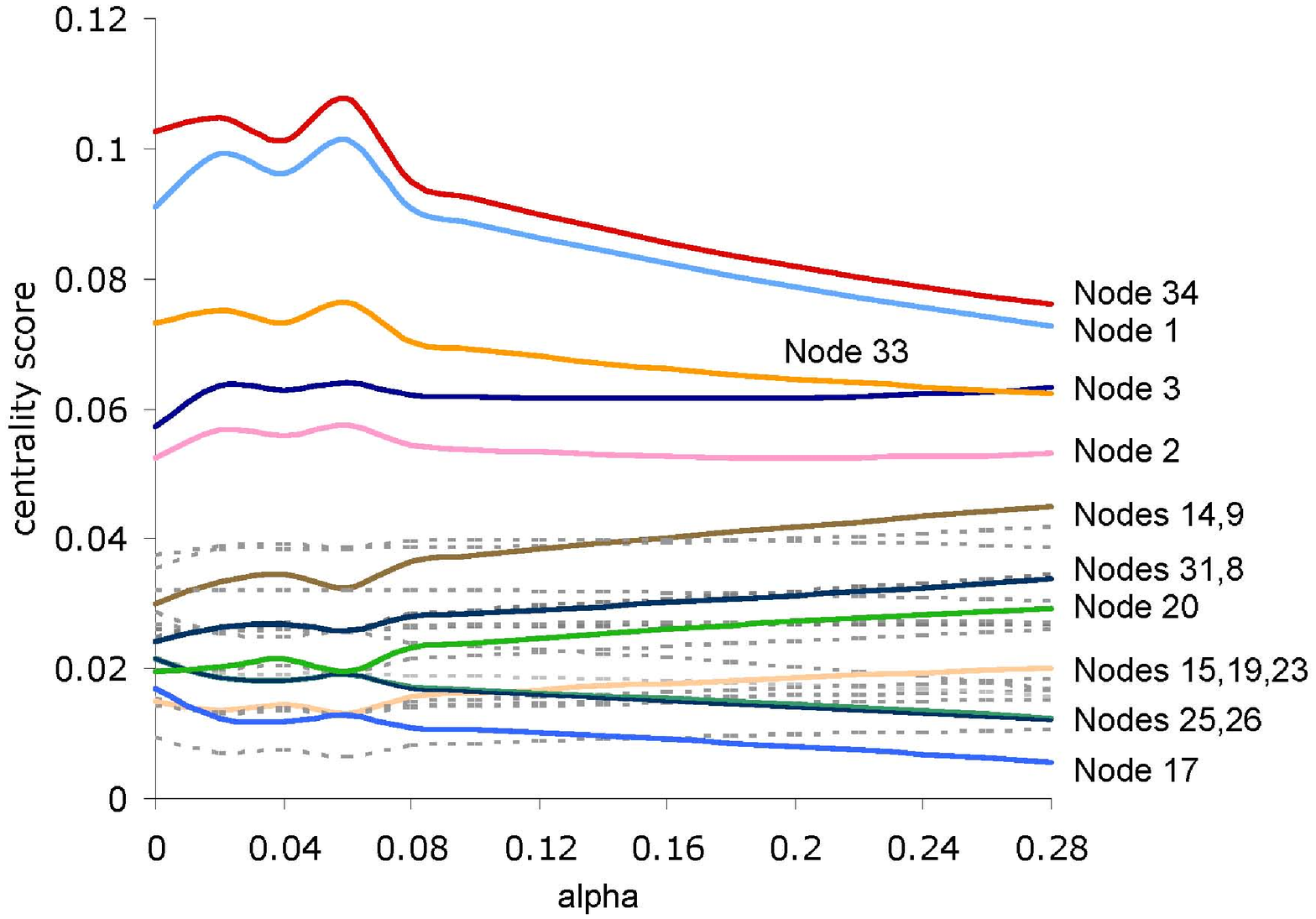} \\
(b)
\end{tabular}
\caption{Zachary's karate club data. (a) Circles and squares represent the two actual factions, while colors stand for discovered communities  for $\alpha=0$. (b) Centrality of club members vs.\ $\alpha$
}
 \label{fig:1}
\end{figure}

First, we study the friendship network of Zachary's karate club~\cite{Zachary}. During the course of the study, a disagreement developed between the administrator and the club's instructor, resulting in the division of the club into two factions, represented by circles and  squares in \figref{fig:1}(a). We find community division of this network for $ 0 \le \alpha \leq 0.29$ (maximum $\alpha$ is given by reciprocal of the largest eigenvalue of the adjacency matrix).

The first bisection of the network results in two communities, regardless of the value of $\alpha$, which are identical to the two factions observed by Zachary.  However, when the algorithm runs to termination (no more bisections are possible), different communities are found for different values of $\alpha$. For $\alpha=0$, the method reduces to edge-based modularity maximization~\cite{Newman204} and leads to four communities (\figref{fig:1}(a)). For  $0<\alpha<0.14$  it discovers three communities, and for $0.14 \le \alpha \le 0.29$,  two communities that are identical to the factions found by Zachary. As Table~\ref{tbl:mi} shows, the purity of discovered communities increases with $\alpha$.

\begin{table}[htdp]
\caption{The number and purity of communities discovered at different values of $\alpha$}
\begin{center}
\begin{tabular}{|c|c|c||c|c|c||c|c|c|}
\hline
\multicolumn{3}{|c||}{\emph{karate club}} &
\multicolumn{3}{c||}{\emph{football}} &
\multicolumn{3}{c|}{\emph{flickr}} \\ \hline
$\alpha$ & grps & Pu & $\alpha$ & grps & Pu & $\alpha$ & grps & Pu\\ \hline
0.00 	& 4 &	0.505 	& 0.00 & 8 	& 0.715	& 0.00    & 4 &	0.501 \\
0.12 	& 3 & 	0.736 	&  0.02 & 8	& 0.723	&  0.001 & 3 &	0.565\\
0.28 	& 2 & 	1.000		& 0.04 & 8	& 0.723	&  0.002 & 3 &	 0.567\\
	& 	&		& 0.06 & 7	& 0.723	&  0.003 & 3 &	0.567\\ \cline{1-3}
\multicolumn{3}{|c||}{\emph{political books}} & 0.08 & 7	& 0.723	&  0.004 & 3 &	0.567\\ \cline{1-3}
0.00 	& 4 &	0.633		& 0.10 & 7	& 0.791	&  0.005 & 3 &	0.568\\
0.04 	& 3 & 	0.805 	& 0.12 & 6	& 0.803	&  0.006 & 3 &	0.570\\
0.08 	& 2 & 	0.917 	&  0.14 & 6	& 0.813	&  0.007 & 3 &	0.571\\
 	&  	&		& 0.16 & 6 	& 0.813 	& 0.008 & 3 &	0.572\\
 	& 	&		& 0.18 & 4	& 0.862	&  0.009 & 3 & 	0.574\\ \hline
\end{tabular}
\end{center}
\label{tbl:mi}
\end{table}

Figure~\ref{fig:1}(b) shows how b-centrality changes with $\alpha$. Nodes 34 and 1\remove{, 33, 2} have  the highest  centrality for all values of $\alpha$. It was the disagreement between these \emph{leaders}, the club administrator (node 1) and instructor (34), that led to the club's division.  Nodes 33 and 2 also have high centrality and hold leadership positions. All these nodes are scored highly by betweenness centrality (BC)~\cite{Freeman} and PageRank (PR)~\cite{PageRank}.
Centrality of nodes 3, 14, 9, 31, 20, 8 increases with $\alpha$ from moderate to relatively high values. All of them (except 8) are connected to both communities: these are the \emph{mediators}. BC scores of these nodes are low, but non-zero. 	 
Nodes 25, 26 and 17 have low centrality which decreases with $\alpha$. These are \emph{peripheral} members. BC of 17 is zero, as expected, but 25 and 26 have scores similar to 31. PR scores of these peripheral nodes are higher than nodes 21, 22, 23 that are connected to central nodes, and comparable to scores of mediator nodes 20 and 31. While both BC and PR correctly pick out leaders, they do not distinguish between peripheral members and mediators.

We also studied the US College football dataset~\cite{GirvanNewman02}
and the political books data\footnote{\texttt{http://www.orgnet.com/}}.
The first network represents the schedule of Division 1 games for the 2001 season where the nodes represent teams and the edges represent the regular season games between teams. The teams are divided into conferences containing 8 to 12 teams each. Games are more frequent between members of the same conference, thought inter-conference games also take place. This leads to an intuition, that the natural communities may be larger than conferences.
The political books network represents books about US politics sold by the online bookseller Amazon. Edges represent frequent co-purchasing  by the same buyers, as indicated by the ``customers who bought this book also bought these other books'' feature of Amazon.  The nodes were labeled \emph{liberal}, \emph{neutral}, or \emph{conservative} by Mark Newman on a reading their descriptions and reviews on Amazon\footnote{\texttt{http://www-personal.umich.edu/$\sim$mejn/netdata/}}.
The number and purity of the communities found in these networks for various values of $\alpha$ are shown in Table~\ref{tbl:mi}.  $\alpha=0$ case corresponds to edge-based modularity method.  As $\alpha$ increases, the number of groups goes down, while their purity increases.
%
We were not able to evaluate rankings due to the lack of gold standard for these datasets.

In addition to benchmark networks, we also studied a social network retrieved from the social photosharing site Flickr. We sampled Flickr's social network by identifying roughly 2000 users interested in one of three topics: \emph{portraiture}, \emph{wildlife}, and \emph{technology}.
Further, we identified four (eight \emph{wildlife}) users who were interested in each topic by studying their profiles, specifically  group membership and user's tags.
We then used Flickr API to retrieve these users' contacts, as well as their contacts' contacts, and labeled all by the topic through which they were discovered.

We reduced the network to an undirected network of mutual contacts only, resulting in a network of $5747$ users, with $1620$, $1337$ and $2790$ users labeled  \emph{technology}, \emph{portraiture} and \emph{wildlife} respectively. Although we did not verify that all the users were interested in the topics they were labeled with, we use these `soft' labels to evaluate the discovered communities. For $\alpha=0$, we found four groups, while for higher values of $\alpha$ ($\alpha<0.01$), we found three groups. As shown in Table~\ref{tbl:mi} the purity of discovered communities increases steadily with $\alpha$.

We can generalize the centrality metric presented above into a notion of path-based connectivity and relate it to other centrality metrics.
Let $q^n_{ij}$ be the number of paths of length $n$ connecting nodes $i$ and $j$. Number of paths of length one connecting $i$ and $j$ is $q^1_{ij}=A_{ij}$, paths of length two is $q^2_{ij}=(A \times A)_{ij}$, etc. The expected number of paths connecting two nodes is:
\begin{equation}
\label {eq:top1}
\mathbf{E(q_{ij})= \frac {(W_{1} \cdot q^1_{ij}+ W_{2} \cdot q^2_{ij}+ \ldots +W_{n} \cdot q^n_{ij}+ \ldots)}{\sum_{i=1}^\infty W_{i}}} \nonumber
\end{equation}
This value can be used to find out how connected two nodes are. Note that $W_{i}$ can be a scalar or a vector.

Several path-based centrality metrics can be expressed in terms of $E(q_{ij})$, including random walk models~\cite{PageRank,Tong06,Tong08,Zhou03,Newman05}, Katz score, as well as Bonacich centrality. In random walk models, a particle starts a random walk at node $i$, and iteratively transitions to its neighbors with probability proportional to the corresponding edge weights. At each step, the particle returns to $i$ with some restart probability ($1-c$). The proximity score is defined as the steady-state probability $r_{i,j}$ that the particle will reach node $j$~\cite{Tong08}.
\begin{itemize}
\item If $W_{k}=c ^{k}\cdot D^{-(k)}$ where $c$ is a constant and  $D$ is  an $n \times n$ matrix with $D_{ij}=\sum_{j=1}^{n} A_{ij} $ if $i=j$ and $0$ otherwise;
then, $E(q_{ij})$ reduces to proximity score in random walk model~\cite{Tong06,Tong08}.

\item If   $W_{i}= \Pi _{j=1}^{i}\alpha_{j}$, where  the scalar $\alpha_{j}$ is the  attenuation factor along the $j$-th link in the path, then $E(q_{ij})$ reduces to Bonacich centrality. For  ease of computation, we have taken  $\alpha_1=\beta$  and $\alpha_i=\alpha$ $\forall i\neq 1$.

\item When $\beta=\alpha$, this  in turn reduces to the Katz status score~\cite{Katz}.

\item When  $\alpha_{1}=1$ and $ \alpha_{2}=\ldots=\alpha_{n}=\ldots=0 $, then $E(q_{ij})$ is the degree centrality used in modularity-maximization approaches~\cite{Newman104}.
\end{itemize}


In summary, we used Bonacich centrality to study the structure of networks, specifically, identify communities and important nodes in the network. We extended the modularity maximization class of algorithms to use b-centrality, rather than edges, as a measure of network connectivity. We applied this approach to  benchmark networks studied in literature and found that it results in network division in close agreement with the ground truth. We also used b-centrality to rank nodes in a network.  By studying changes in rankings that occur when parameter $\alpha$ is varied, we were able to identify locally important `leaders' and globally important `mediators' that facilitate communication between different communities. We can easily extend this definition to multi-modal networks that link entities of different types, and use approach described in this paper to study the structure of complex networks~\cite{Ghosh09socialcom}.

\begin{acknowledgments}
This work is supported in part by the NSF under awards BCS-0527725 and 0915678.
\end{acknowledgments}


\end{document}